\documentclass[preprint,12pt,authoryear]{elsarticle}

\usepackage{amsmath,amssymb,amsfonts}
\usepackage{graphicx}
\usepackage{adjustbox}
\usepackage{booktabs}
\usepackage{multirow}
\usepackage{url}
\usepackage{xcolor}
\usepackage{float}
\usepackage[margin=2cm]{geometry}   
\usepackage{hyperref}               

\journal{Journal of High Energy Astrophysics}

\begin{document}

\begin{frontmatter}

\title{Can next generation CMB experiments discriminate between $\Lambda$CDM and
dynamical dark energy? A Bayesian forecast for Simons Observatory and CMB-S4}

\author[rome]{Anna~Chiara~Ferri\corref{cor1}}
\ead{annachiara.ferri@uniroma1.it}
\author[salamanca]{Ruchika}
\author[rome]{Alessandro~Melchiorri}

\cortext[cor1]{Corresponding author.}

\address[rome]{Physics Department and INFN, Universit\`a di Roma ``La Sapienza'',
Piazzale Aldo Moro 2, 00185, Roma, Italy}
\address[salamanca]{Departamento de Física Fundamental and IUFFyM, Universidad de Salamanca, E-37008 Salamanca, Spain}

\begin{abstract}

The combination of Planck CMB data with BAO measurements from the second data release of DESI has revealed a statistically significant preference for a time evolving dark energy equation of state parameterized in the
Chevallier Polarski Linder (CPL) form. When type Ia supernovae are included, the preference reaches a significance of $4.2\sigma$ with DESY5, $3.8\sigma$ with Union3 and $2.8\sigma$ with Pantheon+. These hint at a departure from $\Lambda$CDM model. 
Our previous Fisher matrix analysis suggested that future CMB experiments could probe such a scenario without external low redshift data sets. Here we extend that study by performing a
full Bayesian MCMC forecast for the Simons Observatory
(SO) and CMB-S4 experiments, assuming as fiducial the DESI+CMB best fit in the
$w_0 w_a$CDM model. We generate mock power spectra with the Cobaya mock CMB pipeline. To quantify the geometric degeneracy in the CMB between $w_0, w_a$ and the matter density, we compute for each mock the
$\Lambda$CDM "shadow" model that preserves the CMB shift parameter $\mathcal{R}$ and produces an equivalent first acoustic peak. SO and CMB-S4 alone cannot distinguish CPL cosmology from its $\Lambda$CDM shadow since the posteriors in the $w_0, w_a$ plane remain broad and largely overlap. This is more pessimistic than our previous Fisher matrix
expectation, showing that the Gaussian approximation underestimates the impact of the geometric degeneracy. Adding BAO and Pantheon$+$ supernovae data, yields  $\sigma(w_0)\simeq 0.05$-$0.06$ for both experiments. However, the resulting preference for dynamical dark energy depends on the BAO data set, for instance, SDSS BAO moves the constraints close to the $\Lambda$CDM. Our results quantify the discriminating power of future CMB surveys in a Bayesian framework, and highlight the role of low redshift distance probes in establishing or rejecting a dynamical dark energy scenario.
\end{abstract}


\end{frontmatter}

\section{Introduction}
\label{sec:introduction}

The concordance $\Lambda$CDM cosmological model, based on the simultaneous presence
of cold dark matter and a cosmological constant $\Lambda$, has been remarkably
successful in describing a very broad range of cosmological observations, from the
cosmic microwave background (CMB) anisotropy pattern measured by the Planck
satellite~\citep{Planck2018} to the large scale clustering of galaxies, to the
distance redshift relation inferred from type Ia supernovae. Nevertheless, with
the continuously increasing precision of cosmological data, a number of
observational tensions have progressively emerged, which challenge the simplest
version of the model~\citep{Abdalla2022,Perivolaropoulos2022,DiValentino2021,
DiValentino2025}. The most prominent is the Hubble tension, i.e. the
$\sim 5\sigma$ discrepancy between the value of $H_0$ inferred from CMB data under
the $\Lambda$CDM assumption and the local determination from the SH0ES
collaboration based on Cepheid calibrated supernovae~\citep{Riess2022,Verde2019}.

In this context, the nature of the component driving the accelerated expansion
of the Universe remains essentially unknown. In the standard cosmological model, dark energy (DE) is described as a perfect fluid with negative pressure, responsible for the observed accelerated expansion of the Universe with an equation of state $w(z) = p_{\rm DE}/\rho_{\rm DE} = -1$. Phenomenologically, however, any small departure from this
value would have important theoretical implications, pointing towards a dynamical
component such as a quintessence scalar field. A commonly adopted parameterization
of a time evolving DE equation of state is the
Chevallier--Polarski--Linder (CPL) form~\citep{Chevallier2001,Linder2003}:
\begin{equation}
w(z) = w_0 + w_a\,\frac{z}{1+z},
\label{eq:cpl}
\end{equation}
where $w_0$ is the equation of state today and $w_a$ parameterizes its evolution
with redshift.

The recent release of the second data (DR2) of the Dark Energy Spectroscopic
Instrument (DESI) collaboration~\citep{DESI_DR2_I,DESI_DR2_II} has provided
unprecedented measurements of the baryon acoustic oscillation (BAO) scale in the
range $0.295 \leq z_{\rm eff} \leq 2.33$. When combined with Planck CMB data and
with one of the three most recent type Ia supernova compilations (Pantheon$+$,
Union3, DES-SN5YR), DESI DR2 BAO measurements point towards a statistically
significant preference for a dynamical dark energy component, reaching the
$2.8\sigma$-$4.2\sigma$ level in the CPL parameterization. The preferred region
in the $w_0, w_a$ plane lies in the quadrant with $w_0 > -1$ today and $w_a < 0$, implying a phantom crossing at intermediate redshifts. A wealth of follow up analyses has confirmed the robustness of this trend across different parameterizations and reconstruction methods
\citep{Lodha2025,Wang2025_NatAstron,Malekjani2024}, while other works have
emphasized that the effect may be partially driven by specific data combinations
\citep{Wang2025_EPJC,Efstathiou2025}.

It is then natural to ask whether the forthcoming generation of CMB experiments
will be able to independently test this scenario, without relying on external
low redshift data. In a previous work~\citep{FerriMelchiorri2025} we addressed
this question using a Fisher matrix forecast, focusing on the Simons Observatory
(SO)~\citep{SO2019} and CMB-S4~\citep{CMBS4_2019}. The conclusion of that study was that, despite the well known geometric degeneracy, the improved sensitivity of future CMB experiments, particularly those approaching the cosmic variance limit, could break this degeneracy sufficiently to discriminate between the DESI preferred $w_0,w_a$ model and $\Lambda$CDM, independently of low redshift observations. However, a Fisher matrix approach
assumes a Gaussian likelihood around the fiducial point and cannot fully capture
the non Gaussianity of the posteriors, in particular the long degeneracy
directions in the $w_0, w_a, H_0, \Omega_m$ parameter space. Furthermore, that analysis accounted only for instrumental noise, neglecting potential systematic uncertainties that could affect the robustness of the forecast constraints.

In the present paper we revisit and extend the analysis of~\citet{FerriMelchiorri2025}
by performing a full Bayesian MCMC forecast for SO, CMB-S4 complemented by an analysis of the actual Planck 2018 data, using as fiducial cosmology the DESI+CMB best fit of the CPL model. We also investigated the constraints on the cosmological parameters obtained by combining these CMB forecasts with low redshift observations, including DESI DR2 and SDSS BAO measurements, as well as Pantheon$+$ Type Ia supernovae data. The paper is organized as follows. In Section~\ref{sec:theory} we review the theoretical motivation and the role of the CMB in constraining dark energy; Section~\ref{sec:method} describes
the methodology, including the mock data generation, the MCMC pipeline and the
$\Lambda$CDM geometric degeneracy construction based on the shift parameter
$\mathcal{R}$; Section~\ref{sec:results} presents the results for SO, CMB-S4 and
Planck, both with CMB alone and in combination with DESI DR2 or SDSS BAO and Pantheon$+$; our conclusions are summarized in Section~\ref{sec:conclusions}.

\section{Theoretical framework and motivation} 
\label{sec:theory}

The CMB anisotropy pattern probes the geometry of the Universe mainly through
the angular scale $\theta_s$ of the sound horizon at recombination, defined as
\begin{equation}
\theta_s \;=\; \frac{r_s(z_\star)}{D_M(z_\star)},
\end{equation}
where $r_s(z_\star)$ is the comoving sound horizon at photon decoupling and
$D_M(z_\star)$ the comoving angular diameter distance to the last scattering
surface at $z_\star \simeq 1090$. The former depends exclusively on pre-recombination
physics (baryon and matter densities, effective number of relativistic species),
while the latter integrates the full expansion history from today back to
decoupling, and is therefore sensitive to the late time ingredients of the model,
including the dark energy sector. What is actually measured from the position of
the first acoustic peak is the combination $\theta_s$; as a consequence, any
change in the late time expansion history can in principle be compensated by an
adjustment of other background parameters, provided $\theta_s$ is left
unchanged. This is exactly the geometric degeneracy of the CMB
\citep{Efstathiou1999,Howlett2012}, which couples $w_0$, $w_a$, $H_0$ and
$\Omega_m$ along an approximately one dimensional manifold in parameter space.

A convenient observable to parameterise this degeneracy is the shift parameter,
\begin{equation}
\mathcal{R} \;=\; \sqrt{\Omega_m H_0^2}\,\frac{D_A(z_\star)\,(1+z_\star)}{c},
\label{eq:shift}
\end{equation}
which encodes the position of the first peak of the temperature angular power
spectrum and is approximately conserved along the degeneracy
direction~\citep{Wang2007,ElgaroyMultamaki2007}. For a given fiducial cosmology in
the CPL model, a $\Lambda$CDM ``shadow'' universe that matches the same CMB peak
position can be constructed by requiring equal values of $\mathcal{R}$ and of
the pre-recombination parameters ($\omega_b, \omega_c, n_s, A_s, \tau$),
leaving $H_0$ free to adjust. We exploit this construction in
Section~\ref{sec:method} to isolate the purely geometric component of the CMB
information on dark energy.

In addition to the geometrical constraint encoded in $\theta_s$, the CMB
provides a second, independent handle on the dark sector through the lensing of its photons by the large scale structure along the line of
sight. The reconstructed CMB lensing potential, together with its imprint on the
temperature and polarisation power spectra via the smoothing of the acoustic
peaks, is sensitive to the late time growth of structure and therefore to any
modification of the dark energy dynamics. This probe has been precisely measured
by Planck~\citep{PlanckLensing2020}, the Atacama Cosmology Telescope (ACT)
\citep{Madhavacheril2024} and the South Pole Telescope~\citep{Ge2025}, and it is
expected to be a key observable for SO and CMB-S4, as well as for the
LiteBIRD~\citep{Lonappan2024} satellite.

The motivation of the present work is therefore twofold. On one hand, we wish to
assess in a fully Bayesian way whether SO and CMB-S4, taken alone, will be able
to discriminate between the CPL best fit preferred by DESI+CMB and a pure
$\Lambda$CDM model. On the other hand, we want to quantify the improvement
brought by the combination with low redshift probes, which is the data combination that originally led to the DESI claim.

\section{Methodology}
\label{sec:method}

\subsection{Fiducial cosmology}

The fiducial cosmology used throughout this paper is taken from the DESI+CMB
MCMC analysis in the $w_0 w_a$CDM model, kindly provided by the authors of
\citet{Giare2024}. The corresponding parameter values and priors adopted in our
MCMC runs are summarized in Table~\ref{tab:fiducial}. The fiducial is obtained in the pure CPL extension of $\Lambda$CDM, with all neutrino species treated as massless$\sum m_\nu = 0 \ \mathrm{eV}$ and an effective number of relativistic species $N_{\rm eff}=3.044$. The same assumption is adopted in all our MCMC runs with the only exception of the Planck analysis where we assume one massive neutrino species with $\sum m_\nu = 0.06 \ \mathrm{eV}$.
The corresponding derived background quantities are $H_0\simeq 63.8$\,km\,s$^{-1}$\,Mpc$^{-1}$ and $\Omega_m\simeq0.349$.

\begin{table}[t]
\centering
\begin{tabular}{lccc}
\toprule
Parameter & Fiducial & Prior range & Prior type \\
\midrule
$\ln(10^{10} A_s)$ & $3.03$ & $[1.61,\,3.91]$ & uniform \\
$n_s$ & $0.9641$ & $[0.8,\,1.2]$ & uniform \\
$100\,\theta_s$ & $1.040693$ & $[0.5,\,10]$ & uniform \\
$\Omega_b h^2$ & $0.02218$ & $[0.018,\,0.028]$ & uniform \\
$\Omega_c h^2$ & $0.11954$ & $[0.08,\,0.18]$ & uniform \\
$w_0$ & $-0.476$ & $[-2.0,\,2.0]$ & uniform \\
$w_a$ & $-1.511$ & $[-3.0,\,3.0]$ & uniform \\
$\tau$ & $0.0501 $ & $0.0501  \pm 0.01$ & Gaussian \\
\bottomrule
\end{tabular}
\caption{Fiducial cosmological parameters used to generate the mock CMB data
and corresponding priors adopted in the MCMC analysis. Flat priors are assumed
unless otherwise indicated. The optical depth $\tau$ is constrained by a
Gaussian prior consistent with current low-$\ell$ Planck data.}
\label{tab:fiducial}
\end{table}

\subsection{Mock CMB data and likelihood} 

To determine how well future CMB observations from the Simons Observatory (SO) and CMB-S4 can constrain the cosmological models considered in this work, we use mock likelihoods.
In particular we use the public \texttt{cobaya\_mock\_cmb} framework\footnote{\url{https://github.com/misharash/cobaya_mock_cmb}}~\citep{CobayaMockCMB}, with experiment specific noise inputs and modifications to include CMB lensing reconstruction. Parameter inference is done with \texttt{cobaya}~\citep{Cobaya2021}, while the theoretical angular power spectra at each sampled cosmology are computed using the Einstein--Boltzmann solver \texttt{CLASS}~\citep{CLASS2011,BlasLesgourguesTram2011}.
For each configuration we run four parallel chains and assess convergence by requiring the Gelman Rubin statistic to satisfy $R-1 < 0.02$. The resulting chains are analyzed with \texttt{GetDist}~\citep{GetDist2019}.

For both experiments, our forecast uses lensed CMB temperature T
and E-mode polarization spectra, $C^{TT}_{\ell}$, $C^{TE}_{\ell}$ and $C^{EE}_{\ell}$, as well as a reconstructed lensing deflection spectrum, $C^{dd}_{L} = L(L + 1)C^{\phi \phi}_{L}$. At each multipole, the theoretical signal-plus-noise covariance of the (T, E) fields is 
\begin{equation}
\boldsymbol{\mathcal C}^{TE}_{\ell} (\theta)  = \begin{pmatrix}
C_{\ell}^{TT}(\theta)+N_{\ell}^{TT} &
C_{\ell}^{TE}(\theta)\\
C_{\ell}^{TE}(\theta) &
C_{\ell}^{EE}(\theta)+N_{\ell}^{EE}
\end{pmatrix},
\end{equation}

where $\theta$ denotes the set of cosmological parameters and $N^{TT}_{\ell}$ and $N^{EE}_{\ell}$ are the experiment specific effective noise spectra. We assume uncorrelated temperature and polarization
noise. The reconstructed lensing field contributes the separate scalar covariance component
\begin{equation}
\mathcal C^{dd}_{L}(\theta) = C^{dd}_{L}(\theta) + N^{dd}_{L},
\end{equation}
where $N_L^{dd}$ is the lensing reconstruction
noise.
We keep the TE correlation but set the correlations of the reconstructed deflection field with temperature and polarization ($C_\ell^{Td}$, $C_\ell^{Ed}$) to zero. In this way the lensing block factorizes from the $(T,E)$ block. The approximation neglects, in particular, the large scale Td correlation and the non Gaussian cross-covariance between the lensed primary spectra and the lensing reconstruction.
The statistical uncertainties associated with the
mock observations arise from cosmic variance,
experimental noise, and the finite sky coverage
of each experiment.
Under the Gaussian field approximation, the covariance between estimated CMB power spectra is given by

\begin{equation}
\operatorname{Cov}
\left(
\widehat C_\ell^{XY},
\widehat C_\ell^{WZ}
\right) =
\frac{
\widetilde C_\ell^{XW}
\widetilde C_\ell^{YZ}
+
\widetilde C_\ell^{XZ}
\widetilde C_\ell^{YW}
}{
(2\ell+1)f_{\mathrm{sky}}
},
\end{equation}

where $\widetilde C_\ell^{XY}
=C_\ell^{XY}+N_\ell^{XY}$ represents the
total signal-plus-noise power spectrum.

In particular, the variance of the CMB power spectrum estimators is

\begin{equation}
\operatorname{Var}(\widehat C_\ell^{XX})
= \frac{2}{(2\ell+1)f_{\mathrm{sky}}}
\left(C_\ell^{XX}+N_\ell^{XX}\right)^2,
\qquad X\in\{T,E\},
\end{equation}

while for the reconstructed lensing deflection
spectrum we have the similar but distinct one

\begin{equation}
\operatorname{Var}(\widehat C_L^{dd})
= \frac{2}{(2L+1)f_{\mathrm{sky}}} \left(C_L^{dd}+N_L^{dd}\right)^2 .
\end{equation}

Here, $f_{\mathrm{sky}}$ represents the observed
sky fraction, which approximately accounts for
the reduction in the number of independent modes
available at each multipole, and differs between experimental setups.
These expressions illustrate how cosmic variance,
experimental noise, and sky coverage determine
the expected statistical uncertainties.
For each experimental configuration, we generate
an idealized mock data set from the theoretical
angular power spectra evaluated at a fixed fiducial
cosmology, $\boldsymbol{\theta}_{\mathrm{fid}}$.

The corresponding fiducial signal-plus-noise
covariances are therefore

\begin{equation}
\widehat{\boldsymbol{\mathcal C}}_\ell^{TE} =
\boldsymbol{\mathcal C}_\ell^{TE}
(\boldsymbol{\theta}_{\mathrm{fid}}) \ , \ \ \ \ \
\widehat{\mathcal C}_L^{dd} = \mathcal C_L^{dd}
(\boldsymbol{\theta}_{\mathrm{fid}}) .
\end{equation}
The mock spectra are therefore noiseless (no realisation scatter is added), so that the posteriors are centred on the fiducial up to projection and prior-volume effects.

During the MCMC sampling, CLASS computes the
theoretical CMB angular power spectra for each
proposed set of cosmological parameters.
These are combined with the corresponding
experimental noise spectra and compared with
the fixed fiducial mock spectra.

The CMB log-likelihood~\citep{HamimecheLewis2008}, normalized
to vanish at the fiducial cosmology, is

\begin{equation}
\begin{aligned}
-2\ln\mathcal L_{\mathrm{CMB}}
={}&
f_{\mathrm{sky}}
\sum_{\ell\in I_{TE}}
(2\ell+1)
\times
\left[
\operatorname{Tr}
\left(
\widehat{\boldsymbol{\mathcal C}}_\ell^{TE}
(\boldsymbol{\mathcal C}_\ell^{TE})^{-1}
\right)
+
\ln
\frac{
\det\boldsymbol{\mathcal C}_\ell^{TE}
}{
\det\widehat{\boldsymbol{\mathcal C}}_\ell^{TE}
}
-2
\right]
\\[6pt]
&+
f_{\mathrm{sky}}
\sum_{L\in I_d}
(2L+1)\times
\left[
\frac{
\widehat{\mathcal C}_L^{dd}
}{
\mathcal C_L^{dd}
}
+
\ln
\frac{
\mathcal C_L^{dd}
}{
\widehat{\mathcal C}_L^{dd}
}
-1
\right].
\end{aligned}
\label{eq:mock_cmb_likelihood}
\end{equation}

The first term represents the joint
temperature polarization Wishart likelihood,
including the correlations between $TT$, $TE$,
and $EE$. The second term describes the independent
lensing deflection contribution within the
adopted Gaussian field approximation.
The multipole sets $I_{TE}$ and $I_d$ are
selected independently for each experimental
configuration, as described below.
No independent $BB$ power spectrum likelihood
is included. Lensing information enters through
both the smoothing of the lensed $TT$, $TE$,
and $EE$ spectra and the reconstructed
deflection field.

\paragraph{Simons Observatory}

For the Simons Observatory (SO) forecasts, we adopt the goal sensitivity configuration of the Large Aperture Telescope implemented through the MockSOGoal likelihood, using the publicly available \texttt{deproj0} temperature and polarization noise curves~\citep{SO2019}. These curves represent the expected post-component-separation sensitivity, including instrumental noise and foreground residuals.
These curves are obtained by combining multi frequency observations through an Internal Linear Combination (ILC) procedure, which minimizes the variance of the resulting CMB maps while preserving the cosmological signal. The deproj0 configuration corresponds to the standard ILC approach, without additional constraints to explicitly remove specific foreground components.
We adopt a sky fraction of $f_{\rm sky}=0.4$ and use the lensed $TT$, $TE$ and $EE$ power spectra over $10\leq\ell\leq4000$.

We also include CMB lensing reconstruction, using the SO minimum variance (MV) reconstruction noise forecasts\footnote{SO lensing noise products: \url{https://github.com/simonsobs/so_noise_models}.}. 
The lensing reconstruction uses multipole ranges that differ from those retained in the primary CMB likelihood,
$30\leq\ell_T\leq3000$ and $30\leq\ell_P\leq5000$.
The published reconstruction noise is provided in terms of the convergence power spectrum $N_L^{\kappa\kappa}$, which we convert to the deflection field convention~\citep{so_noise},
\begin{equation}
N_L^{dd}= L(L+1) N_L^{\phi \phi}=\frac{4N_L^{\kappa\kappa}}{L(L+1)}.
\end{equation}

We add the reconstructed lensing spectrum to the mock likelihood, neglecting its cross-correlation with temperature and polarization. The resulting modified \texttt{MockSOGoal} likelihood includes the lensed primary CMB spectra together with an independent lensing reconstruction contribution.

\paragraph{CMB-S4}

For CMB-S4, we consider the seven year Wide survey of the 2023 conceptual design \citep{CMBS4Design2023,CMBS4_2019} to construct a separate mock likelihood, MockCMBS4PBDR.
We obtain the temperature and polarization noise spectra from the public DRAFT (Dark Radiation Anisotropy Flowdown Team) forecasts \citep{Trendafilova2026}. The DRAFT pipeline models instrumental and atmospheric noise, together with astrophysical foreground emission, and combines the frequency channels through a harmonic space Internal Linear Combination (ILC). We adopt the \texttt{GAL090} sky mask and an effective sky fraction $f_{\rm sky} = 0.62$.
Our forecast includes lensed $TT$, $TE$ and $EE$ spectra for $10\leq\ell\leq5000$, together with the reconstructed CMB lensing deflection spectrum over $10\leq L\leq5000$, using the corresponding minimum variance reconstruction noise projection provided by the
DRAFT pipeline.
As for SO, the published reconstruction noise is expressed in the convergence convention, and is converted into the deflection noise spectrum, and again we neglect the cross-correlation between the primary CMB and the reconstructed lensing field. No independent $BB$ likelihood is used.

\paragraph{Planck}
As a benchmark we also analyse the actual Planck 2018 data, adopting the official binned high-$\ell$ likelihoods together with the low-$\ell$ temperature and polarization likelihoods~\citep{Planck2018Likelihood}. Contrary to the SO and CMB-S4 cases, this is real data and not a mock centred on the CPL fiducial.

\subsection{Construction of the $\Lambda$CDM geometric degeneracy case}

For each of the two CMB configurations we build a $\Lambda$CDM ``shadow''
cosmology that reproduces the same CMB peak position as the $w_0 w_a$CDM
fiducial. Starting from the mean of the posterior of the CPL run, we fix
$\omega_b$, $\omega_c$, $n_s$, $\ln(10^{10} A_s)$, $\tau$ and $100\,\theta_s$ to
the corresponding mean values, set $w_0 = -1$ and $w_a = 0$, and solve for
$H_0$ (and consequently $\Omega_m$). Since $\omega_b$ and $\omega_c$ fix the sound horizon $r_s(z_\star)$, keeping $\theta_s$ fixed is equivalent to keeping $D_M(z_\star)$ and hence the shift parameter
$\mathcal{R}$ of Eq.~(\ref{eq:shift}) unchanged. 
This matches, by construction, the position of the acoustic peaks
and identifies the $\Lambda$CDM point that lies closest to the
$w_0 w_a$CDM fiducial along the geometric degeneracy direction. 
New mock data are generated from this $\Lambda$CDM shadow with the same experimental specifications, and analyzed with a second MCMC run in the same $w_0w_a$CDM parameter space and priors.
The comparison between the two chains for a given experiment quantifies how well that experiment can distinguish, in a Bayesian sense, the CPL fiducial from its closest $\Lambda$CDM counterpart.
The two cosmologies produce nearly indistinguishable first acoustic peaks; any discrimination power can come only from the secondary features of the CMB spectra, in particular from the effects of dark energy on the late time integrated Sachs Wolfe plateau and on the lensing-induced smoothing of the peaks.
To investigate these effects, we consider two experiments with progressively improved sensitivity to CMB lensing: the Simons Observatory (SO), representative of Stage-3 experiments, and CMB-S4, representative of Stage-4 experiments. Although the CMB-S4 project has been discontinued, we retain its forecast specifications as a benchmark to explore the potential of future Stage-4 CMB observations.

\subsection{Combination with DESI DR2, SDSS BAO and Pantheon$+$ Supernovae}

To quantify the impact of low redshift probes, we repeat the analysis by combining the CMB mock likelihoods with BAO measurements from both DESI DR2 \citep{DESI_DR2_II} and SDSS/eBOSS \citep{eBOSS:2020yzd}, as well as the Pantheon$+$ supernova compilation \citep{Scolnic2022,Brout2022}.
In particular, for each CMB experiment, we consider four additional combinations: CMB+DESI, CMB+DESI+Pantheon$+$, CMB+SDSS, and CMB+SDSS+Pantheon$+$. 
In all cases, we adopt the same fiducial CPL cosmology for the CMB mock, while the BAO and SN likelihoods are based on the actual observations. Including both DESI DR2 and SDSS allows us to investigate how the constraints depend on the choice of BAO dataset and to compare the impact of different low-redshift measurements.

These combinations illustrate the constraints that could be obtained by combining future CMB observations with current BAO and supernova data. We stress two caveats. First, the CMB mock is centred on a fiducial that is itself derived from DESI+CMB, so the CMB+DESI combination is not fully independent of the fiducial choice. Second, since the BAO and SN likelihoods are real data, any mismatch between the CPL fiducial and the cosmology preferred by these data shifts the combined posteriors away from the fiducial; the combined results should therefore be read as an assessment of how future CMB data would sharpen current low-redshift constraints, rather than as a self-consistent forecast.

\section{Results}
\label{sec:results}
\subsection{Simons Observatory}
\label{sec:SO}

The posterior constraints obtained with the noiseless Simons Observatory mock for the CPL fiducial, and those of the corresponding $\Lambda$CDM shadow, are reported in Table~\ref{tab:SO}. The recovered mean value of the angular scale of the sound horizon, $100\,\theta_s = 1.04068 \pm 0.00012$, is fully consistent with the input fiducial, confirming the correct calibration of the pipeline. In the CPL case we obtain $w_0 = -0.42 \pm 0.42$ and $w_a = -1.6 \pm 1.0$, with the posterior being broad and strongly elongated along the canonical CMB degeneracy direction. The derived quantities $\Omega_m = 0.368 \pm 0.069$ and $H_0 = 62.8 \pm 5.3$\,km/s/Mpc reflect the
corresponding opening of the degeneracy. 
The CPL constraints are compatible with $\Lambda$CDM: $w_0=-1$ lies at $\sim1.4\sigma$ from the marginalized mean.
When the $\Lambda$CDM shadow ($H_0\simeq68$\,km\,s$^{-1}$\,Mpc$^{-1}$, $\Omega_m\simeq0.31$) is used as fiducial for a second MCMC run with SO noise, the recovered posterior gives $w_0 = -0.92 \pm 0.63$, $w_a = -0.8 \pm 1.3$, $H_0=72\pm12$\,km\,s$^{-1}$\,Mpc$^{-1}$ and $\Omega_m=0.30\pm0.10$, fully compatible with the input. The shift of the posterior mean of $H_0$ above its fiducial value is a projection (prior-volume) effect: when the shadow is analysed in the $w_0w_a$CDM parameter space, the degeneracy extends to large $H_0$ and phantom $w_0$, a region that occupies a large prior volume and pulls the marginalized mean, without any preference of the data for high $H_0$. This should not be interpreted as a tension-reducing effect. For CMB-S4 the tighter lensing constraint truncates this tail, and the mean stays closer to the fiducial ($H_0=64.4\pm8.0$\,km\,s$^{-1}$\,Mpc$^{-1}$).
The two posteriors in the $w_0, w_a$ plane largely overlap, with a separation of $\sim 1\sigma$ along the main degeneracy direction. We note that both posteriors reach the lower edge of the $w_a$ prior, so that the marginalized means and errors on $w_a$ are partially prior dependent.

\begin{table}[H]
\centering
\small
\setlength{\tabcolsep}{4pt}
\begin{tabular}{lcccc}
\toprule
& \multicolumn{2}{c}{SO} & \multicolumn{2}{c}{CMB-S4} \\
\cmidrule(lr){2-3} \cmidrule(lr){4-5}
Parameter & CPL & $\Lambda$CDM shadow & CPL & $\Lambda$CDM shadow \\
\midrule
$100\,\theta_s$
& $1.04068 \pm 0.00012^{\dagger}$
& $1.04070 \pm 0.00012$
& $1.040700 \pm 0.000076^{\dagger}$
& $1.040700 \pm 0.000075$ \\

$\omega_b$
& $0.022162 \pm 0.000056^{\dagger}$
& $0.022165 \pm 0.000055$
& $0.022179 \pm 0.000027^{\dagger}$
& $0.022180 \pm 0.000028$ \\

$\omega_c$
& $0.11955 \pm 0.00082^{\dagger}$
& $0.11927 \pm 0.00080$
& $0.11954 \pm 0.00044^{\dagger}$
& $0.11939 \pm 0.00047$ \\

$\Omega_m$
& $0.368 \pm 0.069$
& $0.30 \pm 0.10$
& $0.359 \pm 0.052$
& $0.357 \pm 0.083$ \\

$H_0$ [km/s/Mpc]
& $62.8 \pm 5.3$
& $72 \pm 12$
& $63.3 \pm 4.4$
& $64.4 \pm 8.0$ \\

$\ln(10^{10} A_s)$
& $3.034 \pm 0.014{\dagger}$ 
& $3.023 \pm 0.017$
& $3.0335 \pm 0.0078{\dagger}$ 
& $3.0312 \pm 0.0082$ \\

$n_s$
& $0.9627 \pm 0.0027^{\dagger}$
& $0.9636 \pm 0.0027$
& $0.9641 \pm 0.0017^{\dagger}$
& $0.9645 \pm 0.0018$ \\

$\tau$
& $0.0509 \pm 0.0076^{\dagger}$
& $0.0459 \pm 0.0088$
& $0.0508 \pm 0.0042^{\dagger}$
& $0.0499 \pm 0.0043$ \\

$w_0$
& $-0.42 \pm 0.42$
& $-0.92 \pm 0.63$
& $-0.41 \pm 0.36$
& $-0.60 \pm 0.60$ \\

$w_a$
& $-1.6 \pm 1.0$
& $-0.8 \pm 1.3$
& $-1.66 \pm 0.87$
& $-1.0 \pm 1.2$ \\

$q_0$
& $0.07 \pm 0.41$
& $-0.56 \pm 0.77$
& $0.08 \pm 0.36$
& $-0.15 \pm 0.67$ \\
\bottomrule
\end{tabular}
\caption{Marginalized means and $68\%$ credible intervals for the SO (left block) and CMB-S4 (right block) forecasts. For each experiment, the first column shows the CPL constraints and the second the associated $\Lambda$CDM shadow, obtained by imposing the same value of the CMB shift parameter $\mathcal{R}$. Parameters kept fixed in the $\Lambda$CDM reconstruction are marked with $^{\dagger}$. Both runs are performed in the $w_0w_a$CDM parameter space.}
\label{tab:SO}
\end{table}

\begin{figure}[H]
\centering
\makebox[\textwidth][c]{%
    \includegraphics[width=0.55\textwidth]{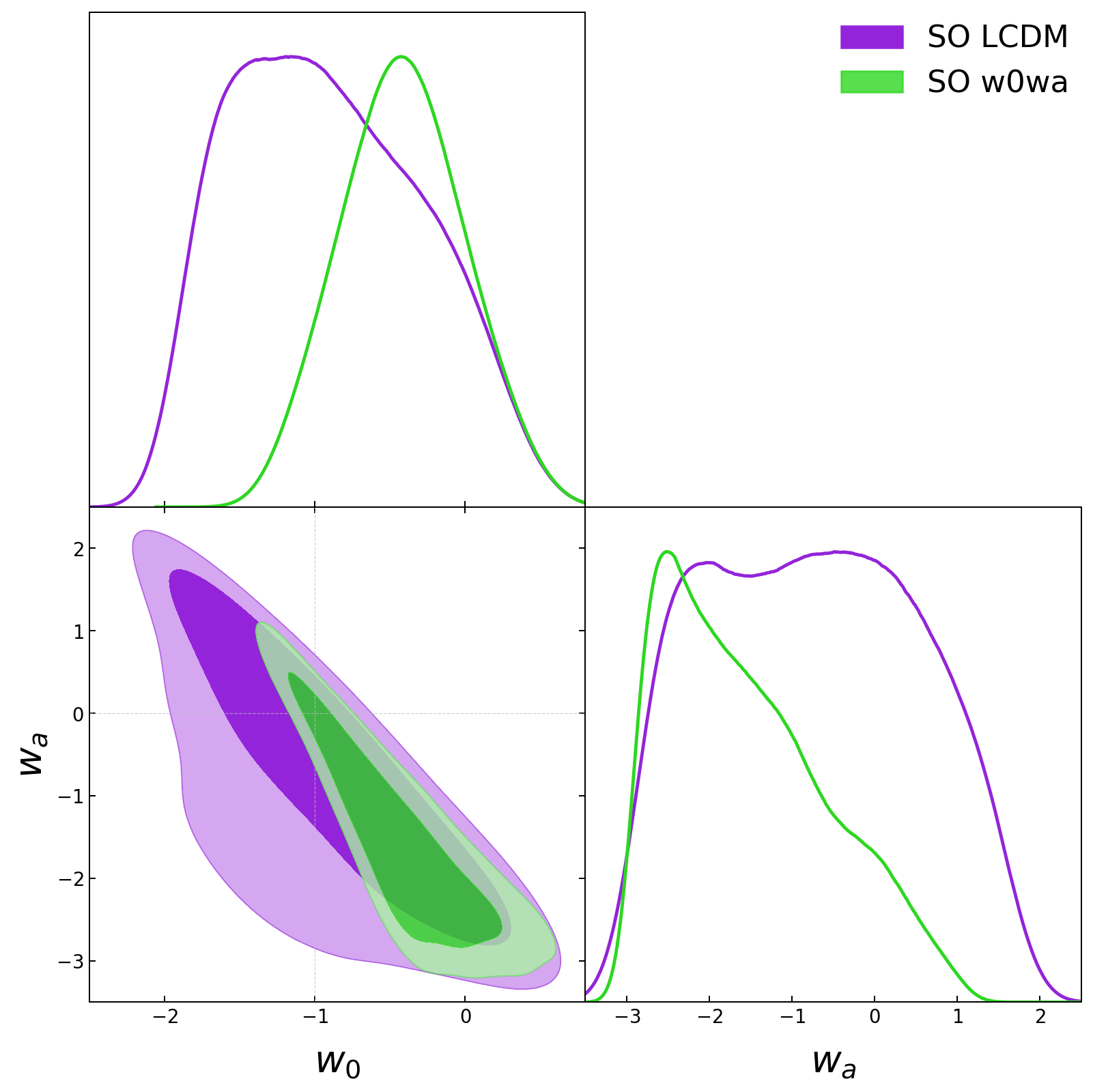}%
    \hspace{0.01\textwidth}%
    \includegraphics[width=0.55\textwidth]{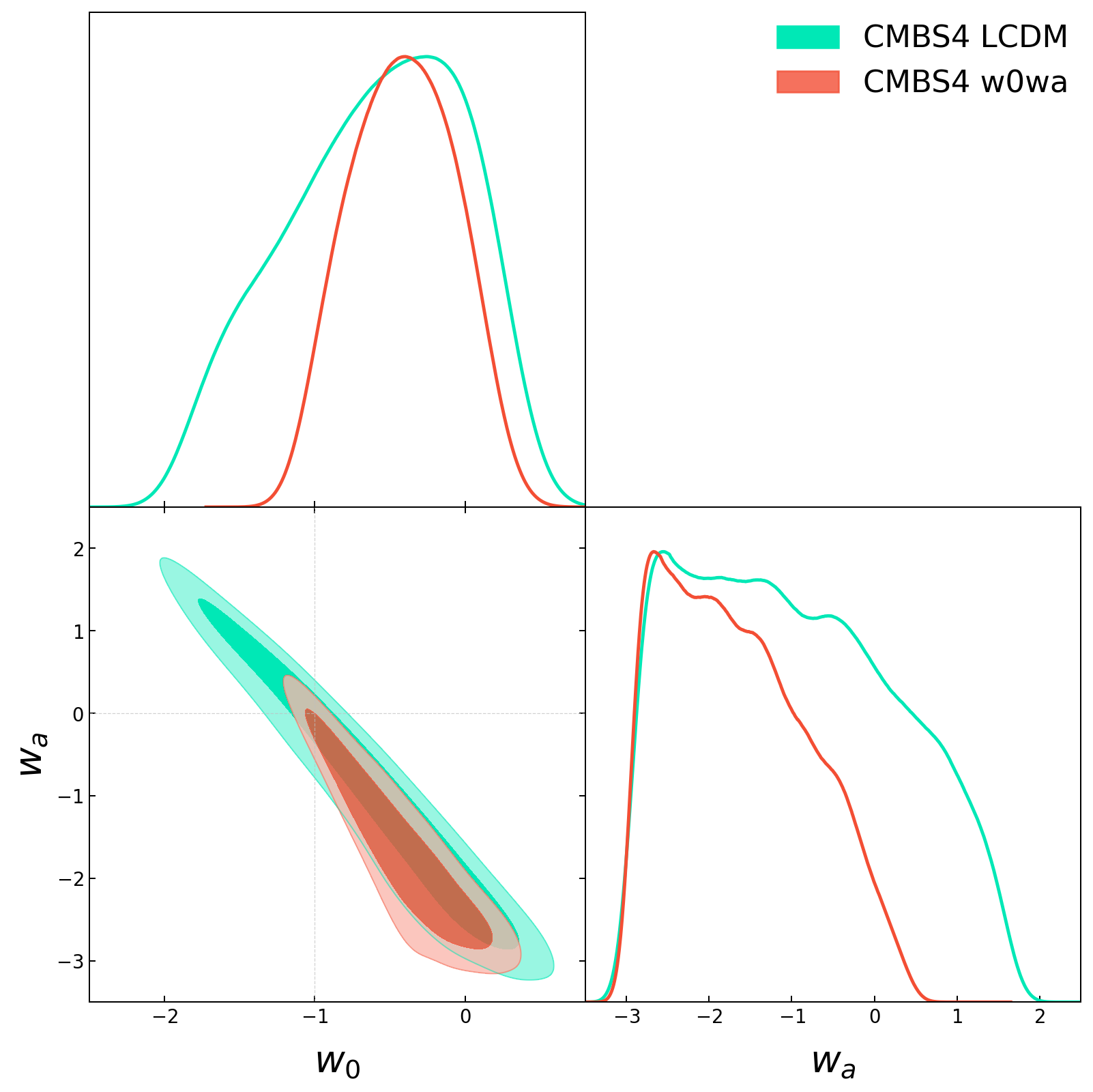}%
}
\caption{Marginalized 68\% and 95\% credible regions in the $(w_0, w_a)$ plane
for the SO (left) and CMB-S4 (right) forecasts, comparing the CPL fiducial
and the associated $\Lambda$CDM shadow. For SO, the two
contours overlap, showing that SO alone cannot discriminate between the two
cosmologies at more than $\sim 1\sigma$. The improvement from CMB-S4 is
modest: the CPL and $\Lambda$CDM contours remain partially overlapping along
the degeneracy direction.}
\label{fig:SO}
\end{figure}

Figure~\ref{fig:SO} shows the marginalized constraints in the $w_0,w_a$ plane obtained from the SO mock likelihood for the CPL fiducial and its associated $\Lambda$CDM shadow.
Although the two models give different marginalized distributions, particularly for $w_0$, their two dimensional credible regions still show a substantial overlap. Both contours are strongly elongated along a direction of correlation between $w_0$ and $w_a$, reflecting the geometric degeneracy of the CMB. Therefore, despite the improved sensitivity of SO compared to Planck, the experiment alone does not provide a clear
separation between the DESI preferred CPL cosmology and its $\Lambda$CDM counterpart.

\subsection{CMB-S4}
\label{sec:CMBS4}

The same analysis is then repeated for CMB-S4. The
corresponding constraints are summarized in Table~\ref{tab:SO} and the $w_0, w_a$ plane is shown in the right panel of Fig.~\ref{fig:SO}. Compared to SO, the increased sensitivity of CMB-S4, in particular to CMB lensing, tightens the constraints, mostly for the CPL fiducial: we find $w_0=-0.41\pm0.36$ and $w_a=-1.66\pm0.87$, a $\sim15\%$ improvement over SO on both parameters. The marginalized posteriors of $w_0$ for the CPL and shadow runs are nearly coincident, while those of $w_a$ differ more visibly, the CPL case favoring more negative values.
The two dimensional credible regions, however, still overlap substantially, even at $68\%$ confidence, and the strong anti correlation between $w_0$ and $w_a$ remains clearly visible, indicating that the geometric degeneracy is not broken by the improved sensitivity.
Even a Stage-4 CMB experiment close to the cosmic variance limit is therefore not sufficient on its own to distinguish the DESI preferred dynamical dark energy model from its associated $\Lambda$CDM shadow.

\subsection{Planck}
\label{sec:Planck}

For completeness, and as a benchmark of current data, we also run the same
$w_0w_a$CDM analysis on the actual Planck 2018 data. The resulting constraints are
summarized in Table~\ref{tab:Planck}. In this case we find
$w_0 =-1.31 \pm 0.49$ and $w_a = -1.2\pm 1.2$, with
$H_0 = 87\pm13$\,km\,s$^{-1}$\,Mpc$^{-1}$. We note that all our analyses are performed using the CLASS definition of the acoustic angular scale, $100\theta_s$. The value $100\theta_s=1.04188\pm0.00028$ reported in Table~\ref{tab:Planck} is obtained from the posterior distribution of the actual Planck data.  The large error bars, and the fact that the Planck posterior opens widely across the prior volume, reflect the inability of Planck alone to constrain simultaneously $w_0$ and
$w_a$ without external data~\citep{Planck2018}.

\begin{table}[H]
\centering
\small
\caption{Posterior means and standard deviations
from the Planck 2018 MCMC analysis in the
CPL $w_0w_a$CDM model.}
\label{tab:Planck}
\begin{tabular}{lc}
\toprule
Parameter & Planck \\
\midrule
$100\,\theta_s$ & $1.04188\pm 0.00028$ \\
$\Omega_b h^2$ & $0.02244 \pm 0.00014$ \\
$\Omega_c h^2$ & $0.1190\pm 0.0012$ \\
$A_s\, [10^{-9}]$ & $2.093\pm0.025$ \\
$n_s$ & $0.9673\pm 0.0040$ \\
$\tau$ & $0.0544 \pm 0.0059$ \\
$w_0$ & $-1.31 \pm 0.49 $ \\
$w_a$ & $-1.2\pm 1.2$ \\
$H_0$ [km/s/Mpc] & $87\pm 13$ \\
$\Omega_m$ & $0.2027\pm 0.070$ \\
$q_0$ & $-1.11\pm 0.66$ \\
\bottomrule
\end{tabular}
\end{table}

\begin{figure}[H]
\centering
\includegraphics[width=0.65\linewidth]{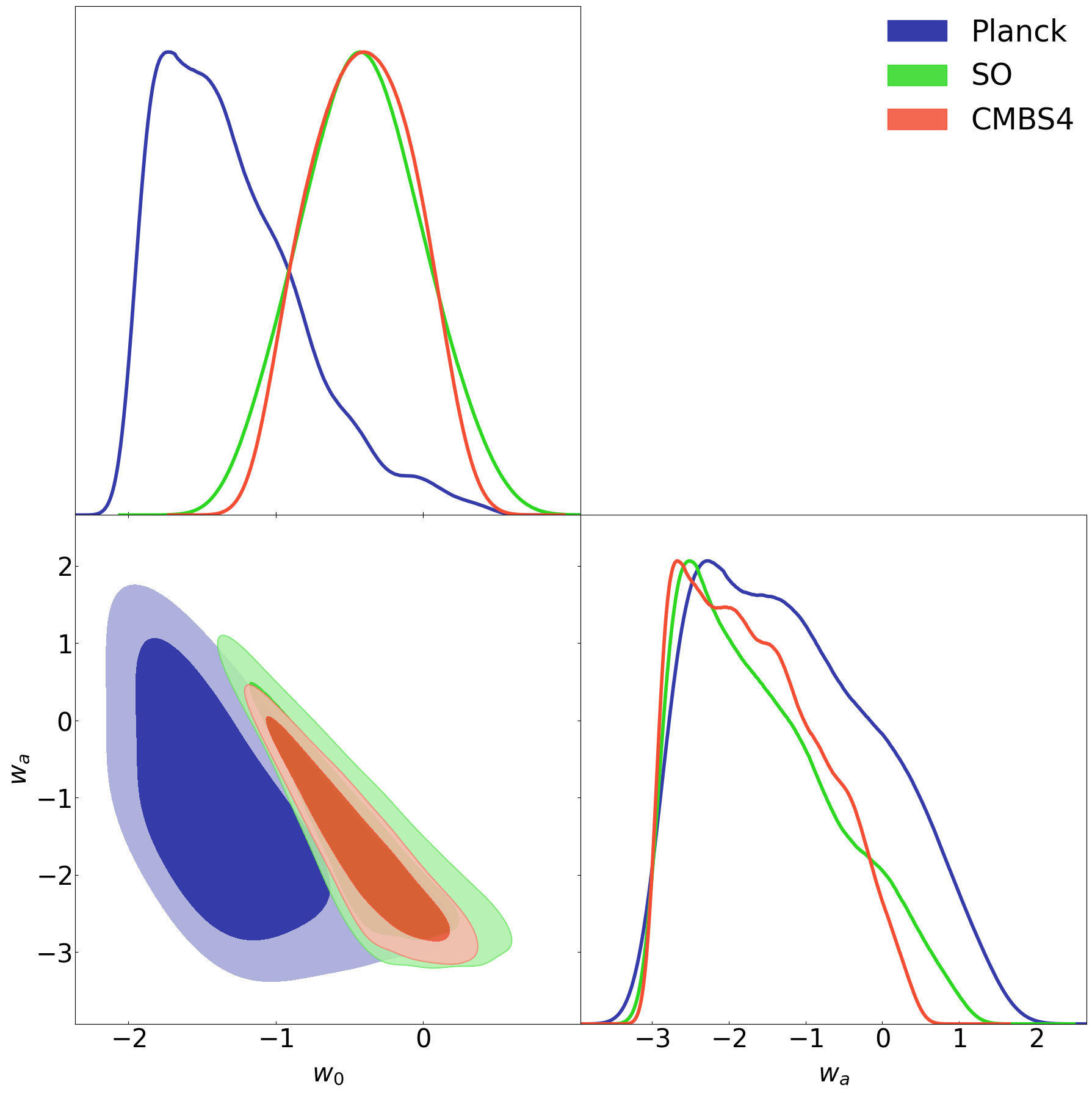}
\caption{Marginalized 68\% and 95\% credible regions in the $w_0, w_a$ plane
for the actual Planck 2018 data and for the SO and CMB-S4 mocks centered on the CPL fiducial. The
improvement brought by SO and CMB-S4 is visible, but the posteriors remain
broad. The Planck contour extends into the phantom region, where the geometric degeneracy opens towards large $H_0$.}
\label{fig:Planck_SO_CMBS4}
\end{figure}

Figure~\ref{fig:Planck_SO_CMBS4} compares the $w_0, w_a$ contours for Planck,
SO and CMB-S4. Since Planck is real data while SO and CMB-S4 are mocks centered on the CPL fiducial, the different positions of the contours are not meaningful by themselves; the relevant comparison is their size. SO and CMB-S4 reduce the allowed parameter
area by roughly a factor of two compared to Planck. Both SO and CMB-S4 contours comfortably include the $\Lambda$CDM value $(w_0, w_a) = (-1, 0)$
inside their $\sim 2\sigma$ regions, confirming that CMB only measurements are
not sufficient to claim a detection of dynamical dark energy.

This result is consistent with the well-known geometric degeneracy of the CMB, which limits the constraints on the dark energy equation of state from CMB observations alone and highlights the importance of complementary low redshift measurements \citep{Efstathiou1999,Howlett2012,Planck2018}.

\subsection{Combination with DESI DR2, SDSS BAO and Pantheon$+$}
\label{sec:combined}

\begin{figure}[H]
\centering
\makebox[\textwidth][c]{%
    \includegraphics[width=0.55\textwidth]{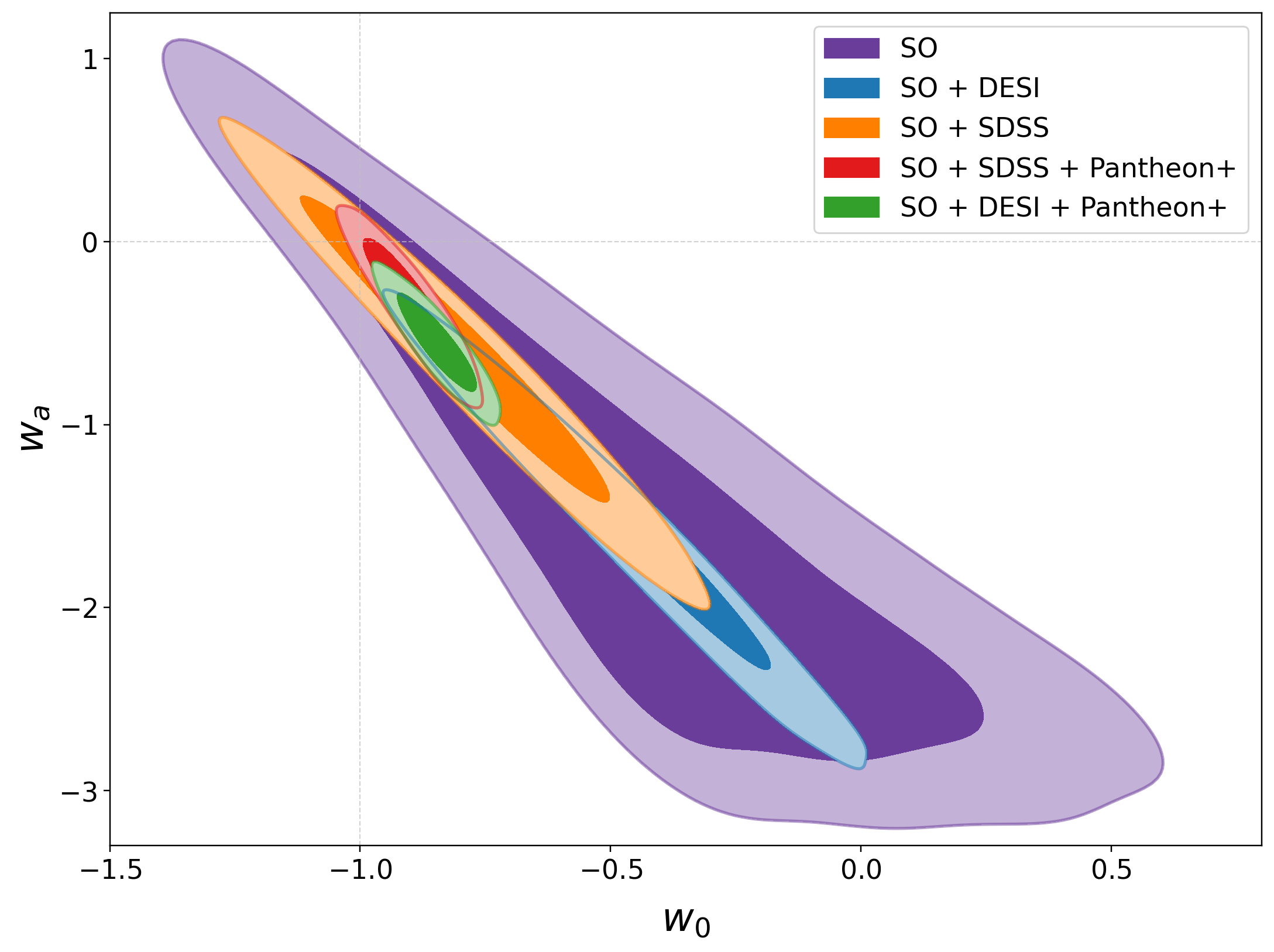}%
    \hspace{0.01\textwidth}%
    \includegraphics[width=0.55\textwidth]{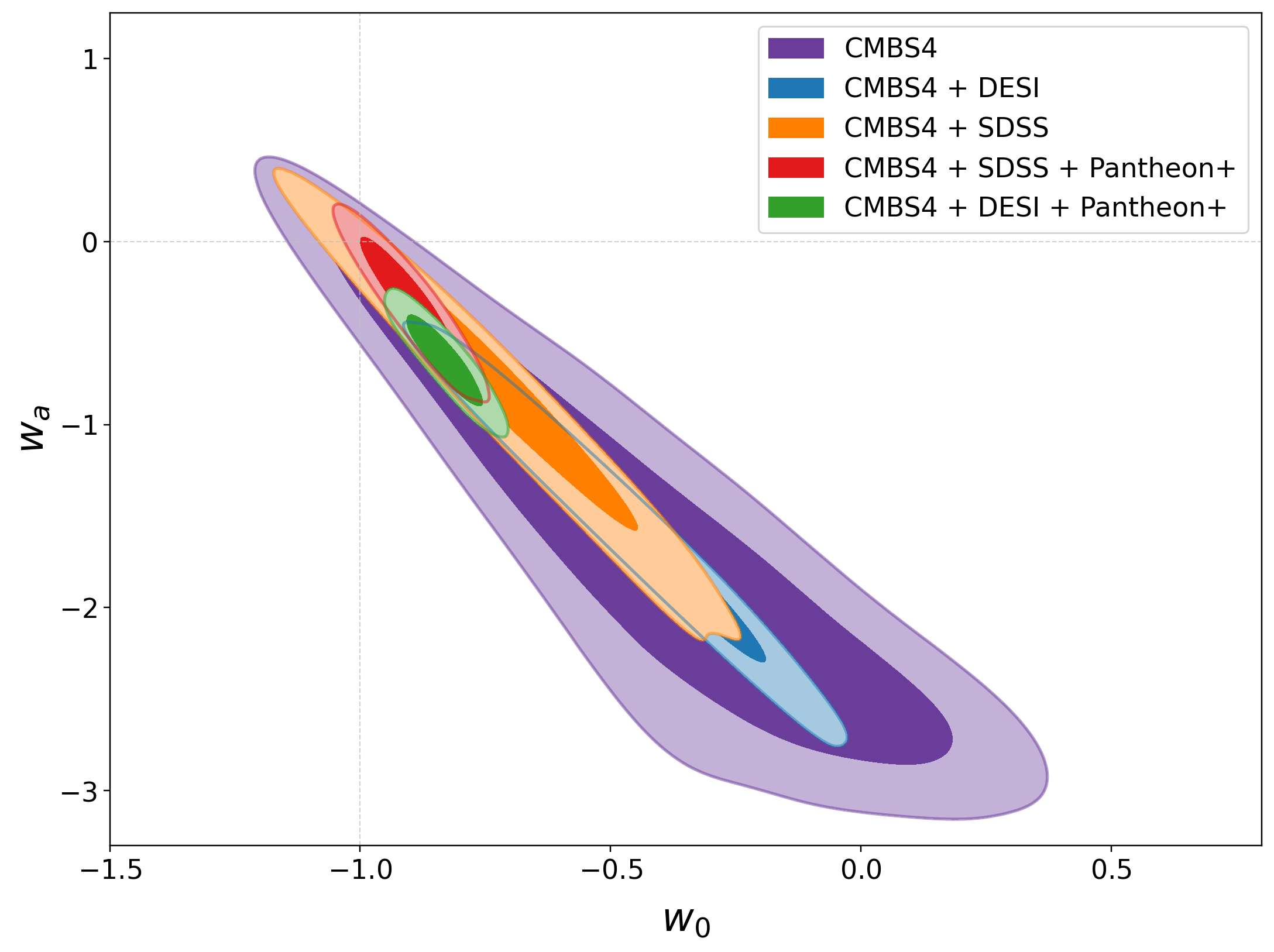}%
}
\caption{ Marginalised 68\% and 95\% credible regions in the $w_0, w_a$ plane
obtained by combining the SO (left) and CMB-S4 (right) mocks with DESI DR2 or SDSS BAO data, with and without Pantheon$+$ supernovae.}
\label{fig:CMB_DESI_SN}
\end{figure}

The picture changes when the mock CMB likelihoods are combined
with actual BAO data and, in addition, with the Pantheon$+$
supernova compilation. Figure~\ref{fig:CMB_DESI_SN} shows the $w_0, w_a$ contours for SO (left) and CMB-S4 (right), alone and combined with DESI DR2, DESI DR2+Pantheon$+$, SDSS, and SDSS+Pantheon$+$. The corresponding constraints are reported in Tables~\ref{tab:SO_mean} and~\ref{tab:CMBS4_mean}.

\begin{table}[H]
\centering
\begingroup
\setlength{\tabcolsep}{2pt}
\renewcommand{\arraystretch}{1.8}

\begin{adjustbox}{max width=\textwidth}
\begin{tabular}{@{}lcccc@{}}
\hline
\hline
Parameter 

& SO + DESI
& SO + DESI + Pantheon$+$
& SO + SDSS
& SO + SDSS + Pantheon$+$ \\
\hline
$100\theta_s$
& $1.0407 \pm 0.00012$
& $1.0407 \pm 0.00012$
& $1.0407 \pm 0.00012$
& $1.0407 \pm 0.00012$ \\

$\omega_b$
& $0.022164 \pm 0.000053$
& $0.022170 \pm 0.000053$
& $0.022159 \pm 0.000050$
& $0.022163 \pm 0.000055$ \\

$\omega_c$
& $0.11935 \pm 0.00065$
& $0.11896 \pm 0.00060$
& $0.11968 \pm 0.00071$
& $0.11963 \pm 0.00071$ \\

$\Omega_m$
& $0.348 \pm 0.021$
& $0.3107 \pm 0.0056$
& $0.325 \pm 0.018$
& $0.3158 \pm 0.0065$ \\

$H_0$ [km/s/Mpc]
& $63.9 \pm 1.9$
& $67.41 \pm 0.60$
& $66.1 \pm 1.9$
& $67.02 \pm 0.66$ \\

$\ln(10^{10} A_s)$
& $3.031 \pm 0.012$
& $3.040 \pm 0.011$
& $3.032 \pm 0.013$
& $3.034 \pm 0.012$ \\

$n_s$
& $0.9633 \pm 0.0024$
& $0.9642 \pm 0.0023$
& $0.9626 \pm 0.0024$
& $0.9626 \pm 0.0024$ \\

$\tau$
& $0.0495 \pm 0.0067$
& $0.0546 \pm 0.0062$
& $0.0497 \pm 0.0073$
& $0.0510 \pm 0.0069$ \\

$w_0$
& $-0.47 \pm 0.20$
& $-0.845 \pm 0.052$
& $-0.79 \pm 0.20$
& $-0.899 \pm 0.060$ \\

$w_a$
& $-1.56 \pm 0.55$
& $-0.56 \pm 0.18$
& $-0.63 \pm 0.55$
& $-0.33 \pm 0.23$ \\

$q_0$
& $0.04 \pm 0.21$
& $-0.373 \pm 0.059$
& $-0.31 \pm 0.22$
& $-0.423 \pm 0.064$ \\
\hline
\hline
\end{tabular}
\end{adjustbox}
\caption{Mean values and standard deviations of the cosmological parameters obtained from SO forecasts combined with different observational datasets.}
\label{tab:SO_mean}
\endgroup
\end{table}


\begin{table}[H]
\centering
\begingroup
\setlength{\tabcolsep}{1pt}
\renewcommand{\arraystretch}{1.8}
\begin{adjustbox}{max width=\textwidth}
\begin{tabular}{@{}lcccc@{}}
\hline
\hline
Parameter
& CMBS4 + DESI
& CMBS4 + DESI + Pantheon$+$
& CMBS4 + SDSS
& CMBS4 + SDSS + Pantheon$+$ \\
\hline

$100\theta_s$
& $1.040700 \pm 0.000076$
& $1.040700 \pm 0.000071$
& $1.040700 \pm 0.000071$
& $1.040700 \pm 0.000076$ \\

$\omega_b$
& $0.022181 \pm 0.000029$
& $0.022177 \pm 0.000026$
& $0.022179 \pm 0.000025$
& $0.022182 \pm 0.000029$ \\

$\omega_c$
& $0.11946 \pm 0.00039$
& $0.11943 \pm 0.00039$
& $0.11973 \pm 0.00040$
& $0.11970 \pm 0.00041$ \\

$\Omega_m$
& $0.348 \pm 0.019$
& $0.3110 \pm 0.0055$
& $0.331 \pm 0.017$
& $0.3161 \pm 0.0062$ \\

$H_0$ [km/s/Mpc]
& $63.8 \pm 1.7$
& $67.49 \pm 0.59$
& $65.6 \pm 1.7$
& $67.00 \pm 0.65$ \\

$\ln(10^{10} A_s)$
& $3.0329 \pm 0.0068$
& $3.0360 \pm 0.0068$
& $3.0335 \pm 0.0075$
& $3.0375 \pm 0.0077$ \\

$n_s$
& $0.9643 \pm 0.0016$
& $0.9644 \pm 0.0015$
& $0.9637 \pm 0.0015$
& $0.9637 \pm 0.0016$ \\

$\tau$
& $0.0507 \pm 0.0039$
& $0.0522 \pm 0.0039$
& $0.0505 \pm 0.0042$
& $0.0525 \pm 0.0043$ \\

$w_0$
& $-0.46 \pm 0.18$
& $-0.830 \pm 0.051$
& $-0.72 \pm 0.19$
& $-0.901 \pm 0.062$ \\

$w_a$
& $-1.57 \pm 0.48$
& $-0.65 \pm 0.16$
& $-0.82 \pm 0.53$
& $-0.32 \pm 0.22$ \\

$q_0$
& $0.04 \pm 0.19$
& $-0.358 \pm 0.057$
& $-0.23 \pm 0.20$
& $-0.425 \pm 0.066$ \\

\hline
\hline
\end{tabular}
\end{adjustbox}
\caption{Mean values and standard deviations of the cosmological parameters obtained from CMB-S4 forecasts combined with different observational datasets.}
\label{tab:CMBS4_mean}
\endgroup
\end{table}
The inclusion of BAO alone already breaks the CMB geometric degeneracy: the allowed region in the $w_0, w_a$ plane collapses along a much tighter linear strip, centered around  the region of parameter space favoring dynamical dark energy, characterized by $w_0>-1$ and $w_a<0$, in agreement with the trends reported by the DESI Collaboration~\citep{DESI_DR2_II,Lodha2025}.
SO+DESI and CMB-S4+DESI yield very similar $1\sigma$ and $2\sigma$ regions.
A similar reduction of the allowed parameter space is observed when combining SO and CMB-S4 with SDSS BAO measurements. However, in this case, the resulting contours are shifted towards the $\Lambda$CDM reference point, $(w_0,w_a)=(-1,0)$, compared to the DESI combinations. This behavior is consistent with the general agreement with $\Lambda$CDM reported in previous analyses involving SDSS/eBOSS measurements~\citep{eBOSS:2020yzd}.
When the Pantheon$+$ supernova sample is further included, the constraints on the dark energy parameters improve significantly for both BAO datasets, with the marginalized uncertainties reaching $\sigma(w_0)\simeq 0.05$/$0.06$ and $\sigma(w_a)\simeq 0.16$/$0.23$.
For both SO and CMB-S4, the combinations with DESI DR2 favor values around $(w_0,w_a)\simeq(-0.84,-0.6)$, with $w_0=-1$ lying at $\sim3\sigma$ from the marginalized mean, while those including SDSS are centered approximately at $(w_0,w_a)\simeq(-0.90,-0.33)$, within $\lesssim 2\sigma$ of the $\Lambda$CDM reference point $(-1,0)$. Note also that, once Pantheon$+$ is included, the combined posteriors move far from the CMB fiducial $(w_0,w_a)=(-0.48,-1.51)$, confirming that in these combinations the location of the contours is set by the low redshift data.
SO and CMB-S4 provide nearly identical constraints once BAO and Pantheon$+$ measurements are included, showing that the constraining power on dynamical dark energy is predominantly driven by the combination with low redshift observations.
These results highlight the importance of combining future CMB observations with BAO and supernova measurements to break the CMB geometric degeneracy. 

\subsection{The low redshift effect on $q_0$}

An interesting result of our analysis concerns the present day deceleration parameter $q_0$. For the CMB only forecasts, we obtain $q_0=0.07\pm0.41$ for SO and $q_0=0.08\pm0.36$ for CMB-S4. Although both distributions have positive mean values, the large uncertainties indicate that CMB observations alone cannot determine the sign of $q_0$, and therefore cannot establish whether the Universe is currently accelerating within the CPL framework. This reflects the limited sensitivity of CMB measurements to the expansion history at very low redshifts, where the present day deceleration parameter is most directly constrained~\citep{Efstathiou1999,Howlett2012,Planck2018}.

The addition of BAO measurements ~\citep{DESI_DR2_II,eBOSS:2020yzd} significantly improves these constraints, although the inferred values depend on the BAO dataset adopted. In particular, the combinations with DESI DR2 retain positive central values of $q_0$, while those with SDSS shift the mean to negative values.
When Pantheon$+$ supernova measurements are further included~\citep{Scolnic2022,Brout2022}, all combinations yield negative values of $q_0$, with uncertainties of approximately $\sigma(q_0)\simeq0.06$, consistently favoring a presently accelerating Universe.

This behavior is also consistent with the findings in ~\citet{FerriRuchikaMelchiorri2026}, where we discussed the importance of low redshift observations in determining the present day cosmic acceleration and showed how differences in the redshift coverage of DESI and SDSS can affect the reconstructed value of $q_0$. 
Our forecasts further illustrate that, even with the improved sensitivity of future CMB experiments, low redshift measurements remain essential for accurately constraining the present day expansion of the Universe.

\section{Conclusions}
\label{sec:conclusions}

In this paper we have presented a Bayesian MCMC forecast for Simons Observatory and CMB-S4, to assess their ability to discriminate between $\Lambda$CDM and the dynamical dark energy scenario currently favored by DESI DR2. 
We have adopted as fiducial cosmology the best fit of the DESI+CMB combination in the Chevallier--Polarski--Linder $w_0 w_a$CDM model, and generated mock CMB temperature, polarization and lensing spectra accordingly.
For each forecast experiment, we have compared the full CPL MCMC run with the $\Lambda$CDM ``shadow'' cosmology sharing the same value of the CMB shift parameter $\mathcal{R}$, which isolates the purely geometric component of the CMB information on dark energy.
We also analysed the actual Planck 2018 data, which show a much broader
posterior with larger excursions of $H_0$ and $\Omega_m$, consistent with the already known inability of Planck alone to constrain $w_0$ and $w_a$ simultaneously.

We conclude that SO and CMB-S4 alone, with mock data generated around the CPL fiducial preferred by DESI results, get broad posteriors in the $w_0, w_a$ plane, largely overlapping with the associated $\Lambda$CDM shadow. 
The discrimination with respect to a pure cosmological constant is only at the $\sim 1\sigma$ level. This is more pessimistic than the Fisher matrix expectations of~\citet{FerriMelchiorri2025}: the Gaussian approximation underlying the Fisher analysis cannot capture the strongly non Gaussian, banana shaped and prior bounded posteriors along the degeneracy direction, and therefore overestimates the discriminating power of CMB only data.
The limiting factor is the geometric degeneracy of the CMB: any
change in the late time expansion history can be compensated by a shift of $H_0$ and $\Omega_m$ while keeping the shift parameter $\mathcal{R}$ fixed.
The lensing information, while present, does not break the degeneracy at a level sufficient to discriminate the CPL fiducial from its $\Lambda$CDM shadow.
The addition of BAO data to the CMB mocks collapses the
degeneracy direction and yields, for both SO and CMB-S4, nearly identical and much tighter contours, with uncertainties further reduced by the inclusion of Pantheon$+$. In the full combination CMB+BAO+SN we find $\sigma(w_0) \simeq 0.05$--$0.06$ and $\sigma(w_a) \simeq 0.16$--$0.23$. With DESI DR2, $w_0=-1$ is disfavored at $\sim3\sigma$, while with SDSS the constraints remain within $\sim2\sigma$ of $\Lambda$CDM: the evidence for dynamical dark energy in these combinations is driven by the choice of BAO data set rather than by the CMB.

Furthermore, our analysis confirms that future CMB observations alone, even at Stage 4 sensitivity, cannot tightly constrain the present day deceleration parameter $q_0$. The inclusion of low redshift probes, particularly Type Ia supernovae, is essential to recover a well constrained negative value of $q_0$, highlighting the importance of the redshift coverage of cosmological observations in determining the present day acceleration of the Universe~\citep{FerriRuchikaMelchiorri2026}.

Several extensions of the present analysis are worth pursuing. First, the inclusion of the delensed $EE$ power spectrum following the method of~\citet{Ge2025} and of the full lensing--primary cross-covariance could partially alleviate the geometric degeneracy. Second, a self-consistent forecast should combine the future CMB data with mock DESI and supernova data generated at the same fiducial. Third, the addition of
galaxy lensing cross correlations (e.g. with Euclid or LSST) and of kinetic Sunyaev-Zel'dovich measurements may bring additional constraining power on the late time growth of structure and should be investigated in a joint forecasting framework. Finally, an analogous study performed in alternative parameterizations beyond CPL, such as the Barboza Alcaniz or Pad\'e forms~\citep{Malekjani2024}, or using a model independent reconstruction~\citep{Lodha2025}, would help disentangle genuine deviations from $\Lambda$CDM from possible artifacts of the CPL parameterization.

\section*{CRediT authorship contribution statement}
\textbf{Anna Chiara Ferri:} Conceptualisation, Methodology, Software, Formal
analysis, Data curation, Visualisation, Writing -- original draft.\\
\textbf{Ruchika:} Conceptualisation, Methodology, Supervision,
Writing -- review \& editing.\\
\textbf{Alessandro Melchiorri:} Conceptualisation, Methodology, Supervision,
Writing -- review \& editing.

\section*{Declaration of competing interest}
The authors declare that they have no known competing financial interests or
personal relationships that could have appeared to influence the work
reported in this paper.

\section*{Data availability}
The MCMC chains and the numerical products used in this work are available
from the authors upon reasonable request.

\section*{Acknowledgements}
The authors thank Jesús Torrado for initial discussions regarding the Cobaya setup. The authors also acknowledge the use of computational resources provided by CINECA through the ISCRA project IscrC\_CMB4DE.
A.C.F.\ and A.M.\ thank TASP, iniziativa specifica INFN, for financial
support. This work was partially supported by the research grant n.\
2022E2J4RK ``PANTHEON: Perspectives in Astroparticle and Neutrino THEory
with Old and New messengers'' under the PRIN 2022 programme funded by the
Italian Ministero dell'Universit\`a e della Ricerca (MUR). A.C.F. would also like to thank the University of Salamanca for its hospitality during a research visit, where part of this work was carried out.
R.\ acknowledges financial support from Grant PID2024-158938NB-I00, funded by MICIU/AEI/10.13039/501100011033 and co-funded by ERDF/EU (``A way of making Europe''), as well as from Project SA097P24, funded by the Junta de Castilla y León.


\begin{thebibliography}{99}
\bibitem[Abdalla et al.(2022)]{Abdalla2022}
E.~Abdalla et al.,
\newblock Cosmology intertwined: A review of the particle physics, astrophysics,
and cosmology associated with the cosmological tensions and anomalies,
\newblock \textit{JHEAp} 34 (2022) 49.

\bibitem[Alam et al.(2021)]{eBOSS:2020yzd}
S.~Alam et al. (eBOSS Collaboration),
\newblock Completed SDSS-IV extended Baryon Oscillation Spectroscopic Survey:
Cosmological implications from two decades of spectroscopic surveys at the
Apache Point Observatory,
\newblock \textit{Phys. Rev. D} 103 (2021) 083533,
arXiv:2007.08991.

\bibitem[Blas et al.(2011)]{BlasLesgourguesTram2011}
D.~Blas, J.~Lesgourgues and T.~Tram,
\newblock The Cosmic Linear Anisotropy Solving System (CLASS).
Part II: Approximation Schemes,
\newblock \textit{JCAP} 07 (2011) 034,
arXiv:1104.2933.

\bibitem[Brout et al.(2022)]{Brout2022}
D.~Brout et al.,
\newblock The Pantheon$+$ analysis: cosmological constraints,
\newblock \textit{Astrophys. J.} 938 (2022) 110.

\bibitem[Chevallier and Polarski(2001)]{Chevallier2001}
M.~Chevallier and D.~Polarski,
\newblock Accelerating universes with scaling dark matter,
\newblock \textit{Int. J. Mod. Phys. D} 10 (2001) 213.

\bibitem[CMB-S4 Collaboration(2019)]{CMBS4_2019}
K.~Abazajian et al. (CMB-S4 Collaboration),
\newblock CMB-S4 Science Case, Reference Design, and Project Plan,
\newblock arXiv:1907.04473 (2019).

\bibitem[CMB-S4 Collaboration(2023)]{CMBS4Design2023}
CMB-S4 Collaboration,
\newblock CMB-S4 Design Report 2023,
\newblock CMB-S4 Technical Report, CMBS4-doc-716-v6 (2023).

\bibitem[DESI Collaboration(2025a)]{DESI_DR2_I}
M.~Abdul Karim et al. (DESI Collaboration),
\newblock DESI DR2 results. I. Baryon acoustic oscillations from the Lyman-alpha forest,
\newblock \textit{Phys. Rev. D} 112 (2025) 083514.

\bibitem[DESI Collaboration(2025b)]{DESI_DR2_II}
M.~Abdul Karim et al. (DESI Collaboration),
\newblock DESI DR2 results. II. Measurements of baryon acoustic oscillations
and cosmological constraints,
\newblock \textit{Phys. Rev. D} 112 (2025) 083515.

\bibitem[Di Valentino et al.(2021)]{DiValentino2021}
E.~Di Valentino et al.,
\newblock In the realm of the Hubble tension -- a review of solutions,
\newblock \textit{Class. Quantum Grav.} 38 (2021) 153001.

\bibitem[Di Valentino et al.(2025)]{DiValentino2025}
E.~Di Valentino et al.,
\newblock The CosmoVerse White Paper: addressing observational tensions in
cosmology with systematics and fundamental physics,
\newblock arXiv:2504.01669 (2025).

\bibitem[Efstathiou and Bond(1999)]{Efstathiou1999}
G.~Efstathiou and J.~R.~Bond,
\newblock Cosmic confusion: degeneracies among cosmological parameters derived
from measurements of microwave background anisotropies,
\newblock \textit{Mon. Not. R. Astron. Soc.} 304 (1999) 75--97,
arXiv:astro-ph/9807103.

\bibitem[Efstathiou(2025)]{Efstathiou2025}
G.~Efstathiou,
\newblock Evolving dark energy or supernovae systematics?,
\newblock \textit{Mon. Not. R. Astron. Soc.} 538 (2025) 875.

\bibitem[Elgar\o y and Multam\"aki(2007)]{ElgaroyMultamaki2007}
{\O}.~Elgar{\o}y and T.~Multam\"aki,
\newblock On using the CMB shift parameter in tests of models of dark energy,
\newblock \textit{Astron. Astrophys.} 471 (2007) 65.

\bibitem[Ferri and Melchiorri(2026)]{FerriMelchiorri2025}
A.C.~Ferri and A.~Melchiorri,
\newblock Can future CMB data discriminate between a cosmological constant and
dynamical dark energy?,
\newblock \textit{JHEAp} 50 (2026) 100504,
doi:10.1016/j.jheap.2025.100504.

\bibitem[Ferri et al.(2026)]{FerriRuchikaMelchiorri2026}
A.C.~Ferri, Ruchika and A.~Melchiorri,
\newblock Present Day Cosmic Acceleration from SDSS and DESI BAO:
A Call for Finer Tomography of the DESI Bright Galaxy Survey,
\newblock arXiv:2607.07348 (2026).

\bibitem[Ge et al.(2025)]{Ge2025}
F.~Ge et al.,
\newblock Cosmology from CMB lensing and delensed EE power spectra using
2019--2020 SPT-3G polarization data,
\newblock \textit{Phys. Rev. D} 111 (2025) 083534.

\bibitem[Giar\`e et al.(2024)]{Giare2024}
W.~Giar\`e, M.~Najafi, S.~Pan, E.~Di Valentino, and J.T.~Firouzjaee,
\newblock Robust preference for dynamical dark energy in DESI BAO and SN
measurements,
\newblock \textit{JCAP} 10 (2024) 035.

\bibitem[Gu et al.(2025)]{Wang2025_NatAstron}
G.~Gu et al.,
\newblock Dynamical dark energy in light of the DESI DR2 baryonic acoustic
oscillations measurements,
\newblock \textit{Nat. Astron.} 9 (2025) 1879--1889,
doi:10.1038/s41550-025-02669-6.

\bibitem[Hamimeche and Lewis(2008)]{HamimecheLewis2008}
S.~Hamimeche and A.~Lewis,
\newblock Likelihood analysis of CMB temperature and polarization power spectra,
\newblock \textit{Phys. Rev. D} 77 (2008) 103013,
arXiv:0801.0554.

\bibitem[Howlett et al.(2012)]{Howlett2012}
C.~Howlett, A.~Lewis, A.~Hall and A.~Challinor,
\newblock CMB power spectrum parameter degeneracies in the era of precision
cosmology,
\newblock \textit{JCAP} 04 (2012) 027,
arXiv:1201.3654.

\bibitem[Lesgourgues(2011)]{CLASS2011}
J.~Lesgourgues,
\newblock The Cosmic Linear Anisotropy Solving System (CLASS) I: Overview,
\newblock arXiv:1104.2932 (2011).

\bibitem[Lewis(2019)]{GetDist2019}
A.~Lewis,
\newblock GetDist: a Python package for analysing Monte Carlo samples,
\newblock arXiv:1910.13970 (2019).

\bibitem[Linder(2003)]{Linder2003}
E.V.~Linder,
\newblock Exploring the expansion history of the universe,
\newblock \textit{Phys. Rev. Lett.} 90 (2003) 091301.

\bibitem[Lodha et al.(2025)]{Lodha2025}
K.~Lodha et al. (DESI Collaboration),
\newblock Extended dark energy analysis using DESI DR2 BAO measurements,
\newblock \textit{Phys. Rev. D} 112 (2025) 083511.

\bibitem[Lonappan et al.(2024)]{Lonappan2024}
A.I.~Lonappan et al.,
\newblock LiteBIRD science goals and forecasts: a full-sky measurement of
gravitational lensing of the CMB,
\newblock \textit{JCAP} 06 (2024) 009.

\bibitem[Madhavacheril et al.(2024)]{Madhavacheril2024}
M.S.~Madhavacheril et al. (ACT Collaboration),
\newblock The Atacama Cosmology Telescope: DR6 gravitational lensing map and
cosmological parameters,
\newblock \textit{Astrophys. J.} 962 (2024) 113.

\bibitem[Malekjani et al.(2024)]{Malekjani2024}
M.~Malekjani, R.~Md Akhlaghi, and M.~Rezaei,
\newblock Cosmological constraints on dark energy parametrizations after
DESI 2024: persistent deviation from standard $\Lambda$CDM cosmology,
\newblock arXiv:2407.09767 (2024).

\bibitem[Perivolaropoulos and Skara(2022)]{Perivolaropoulos2022}
L.~Perivolaropoulos and F.~Skara,
\newblock Challenges for $\Lambda$CDM: an update,
\newblock \textit{New Astron. Rev.} 95 (2022) 101659.

\bibitem[Planck Collaboration(2020a)]{Planck2018Likelihood}
N.~Aghanim et al. (Planck Collaboration),
\newblock Planck 2018 results. V. CMB power spectra and likelihoods,
\newblock \textit{Astron. Astrophys.} 641 (2020) A5.

\bibitem[Planck Collaboration(2020b)]{Planck2018}
N.~Aghanim et al. (Planck Collaboration),
\newblock Planck 2018 results. VI. Cosmological parameters,
\newblock \textit{Astron. Astrophys.} 641 (2020) A6,
arXiv:1807.06209.

\bibitem[Planck Collaboration(2020c)]{PlanckLensing2020}
N.~Aghanim et al. (Planck Collaboration),
\newblock Planck 2018 results. VIII. Gravitational lensing,
\newblock \textit{Astron. Astrophys.} 641 (2020) A8.

\bibitem[Rashkovetskyi(2024)]{CobayaMockCMB}
M.~Rashkovetskyi,
\newblock \texttt{cobaya\_mock\_cmb}: a mock CMB likelihood package for Cobaya,
\newblock \url{https://github.com/misharash/cobaya_mock_cmb} (2024).

\bibitem[Riess et al.(2022)]{Riess2022}
A.G.~Riess et al.,
\newblock A comprehensive measurement of the local value of the Hubble
constant with 1 km\,s$^{-1}$\,Mpc$^{-1}$ uncertainty from the Hubble Space
Telescope and the SH0ES team,
\newblock \textit{Astrophys. J. Lett.} 934 (2022) L7.

\bibitem[Scolnic et al.(2022)]{Scolnic2022}
D.~Scolnic et al.,
\newblock The Pantheon$+$ analysis: the full data set and light-curve release,
\newblock \textit{Astrophys. J.} 938 (2022) 113.

\bibitem[Simons Observatory Collaboration(2019)]{SO2019}
P.~Ade et al. (Simons Observatory Collaboration),
\newblock The Simons Observatory: science goals and forecasts,
\newblock \textit{JCAP} 02 (2019) 056,
arXiv:1808.07445.

\bibitem[Simons Observatory Collaboration(2026)]{so_noise}
Simons Observatory Collaboration,
\newblock SO noise models and lensing reconstruction forecasts,
\newblock \url{https://github.com/simonsobs/so_noise_models}
(accessed October 2026).

\bibitem[Torrado and Lewis(2021)]{Cobaya2021}
J.~Torrado and A.~Lewis,
\newblock Cobaya: code for Bayesian analysis of hierarchical physical models,
\newblock \textit{JCAP} 05 (2021) 057.

\bibitem[Trendafilova et al.(2026)]{Trendafilova2026}
C.~Trendafilova, S.~Raghunathan, B.~Wallisch, J.~Meyers et al.,
\newblock Sensitivity of Next-Generation CMB Surveys to Neutrinos and Other Light Relics,
\newblock arXiv:2608.07453 (2026).

\bibitem[Verde et al.(2019)]{Verde2019}
L.~Verde, T.~Treu, and A.G.~Riess,
\newblock Tensions between the early and the late Universe,
\newblock \textit{Nat. Astron.} 3 (2019) 891.

\bibitem[Wang and Mota(2025)]{Wang2025_EPJC}
D.~Wang and D.~Mota,
\newblock Did DESI DR2 truly reveal dynamical dark energy?,
\newblock \textit{Eur. Phys. J. C} 85 (2025) 1356.

\bibitem[Wang and Mukherjee(2007)]{Wang2007}
Y.~Wang and P.~Mukherjee,
\newblock Observational constraints on dark energy and cosmic curvature,
\newblock \textit{Phys. Rev. D} 76 (2007) 103533.

\end{thebibliography}
\end{document}